\documentclass[prc,aps,showpacs,floatfix,twocolumn,groupedaddress,amsmath,amssymb]{revtex4}
\usepackage{graphicx}
\usepackage{dcolumn}
\usepackage{bm}
\usepackage{epsfig}
\usepackage{xcolor}

\newcommand{\be}{\begin{equation}}
\newcommand{\ee}{\end{equation}}
\newcommand{\beq}{\begin{eqnarray}}
\newcommand{\eeq}{\end{eqnarray}}
\begin{document}

\title{Possible Shape Coexistence and 
 Magnetic Dipole Transitions in $^{17}$C and $^{21}$Ne}

\author{H.  Sagawa~$^{\rm a}$,  
   ~X. R. Zhou~$^{\rm b}$,  Toshio Suzuki$^{\rm c}$, and  N. Yoshida $^{\rm d}$,}
\address{$^{\rm a}$~Center for Mathematics and Physics, the University of Aizu \\
Aizu-Wakamatsu, Fukushima 965-8580,  Japan\\
$^{\rm b}$~Department of Physics and Institute of Theoretical Physics and Astrophysics \\
 Xiamen University, Xiamen 361005, People's Republic of China\\
$^{\rm c}$~ Department of Physics, College of Humanities and
 Sciences, Nihon University\\     
Sakurajosui 3-25-40, Setagaya-ku, Tokyo 156-8550, Japan\\
$^{\rm d}$~Faculty of Informatics, Kansai University, Takatsuki 
  569-1095, Japan}


\begin{abstract}
Magnetic dipole(M1) transitions of N=11 nuclei, 
 $^{17}$C  and $^{21}$Ne  are investigated by using shell model and deformed 
Skyrme Hartree-Fock+blocked BCS  wave functions. 
  Shell model calculations  predict well 
observed 
 energy spectra and magnetic dipole transitions in  $^{21}$Ne, while 
the results are   rather poor to predict these observables in  $^{17}$C.
In the deformed HF calculations, the ground states of  two nuclei 
are shown to have large prolate deformations close to $\beta_2$=0.4.
It is also pointed out that the first $K^{\pi}=1/2^+$ state
 in  $^{21}$Ne is prolately deformed, while the first $K^{\pi}=1/2^+$ state
 in   $^{17}$C is predicted to have a large 
oblate deformation being  
  close to the ground state in energy, 
We point out  that experimentally observed 
 large hindrance  of M1 transition between $I^{\pi}=1/2^+$ and $3/2^+$
 in  $^{17}$C can  be attributed to 
 a shape coexistence near the ground state of  $^{17}$C.
 \end{abstract}
\maketitle

\section{Introduction}
Recently, many experimental and theoretical efforts have been 
paid to study structure and reaction mechanism in nuclei near drip lines.
 It has been 
known that  
electromagnetic observables  give useful information to study 
structure of nuclei, not only  ground states but also excited states.
These observables are expected to pin down precise 
information of  the configuration and the deformation in  nuclei.
Advanced experimental instruments reveal several unexpected 
  structure of light nuclei with the mass number A$\sim$(10-20).
 One of the current issues 
 is  a large quenching of 
   magnetic dipole  (M1) transition
between the first excited 1/2$^+$ state and the ground state with
 $I^{\pi}$=3/2$^+$ in $^{17}$C \cite{RIKENC17} 
in comparison with the corresponding 
 transitions in one of N=11 isotones, $^{21}$Ne \cite{Fire}.
  
The deformation manifests itself in observables like 
   E2 and M1 moments.  
   In ref.~\cite{Sagawa04}, deformed Skyrme HF+BCS calculations
were performed to study the evolution of deformations  
  in C and Ne isotopes 
 The calculated electric quadrupole moments and magnetic moments
were successfully  compared with empirical data.
 It was pointed out that the shell occupancy 
gives the crucial effect on the evolution of
 the deformation of isotope chains. 
  This  deformation driving mechanism due to the shell occupancy has been  
 noticed as the nuclear Jahn-Teller effect~\cite{NJT},  which gives 
 an intuitive understanding for the evolution of deformation. 
 A  possible shape 
coexistence is pointed out   in $^{17}$C 
because of   different deformation driving effects between  
neutrons  and protons.  Namely, the first excited 
 $K^{\pi}=1/2^+$ state has oblate deformation and 
 almost degenerate with the prolately 
deformed ground state with  $K^{\pi}=3/2^+$. 
On the other hand, there is no sign of shape coexistence in $^{21}$Ne 
   since the 
shell occupancies are almost the same between protons and neutrons. 
In theoretical 
point of view, it is  interesting to see how much differences and 
similarities will appear between the results of standard shell model
 calculations
  and those of the mean field theories.
To this end, the  HF results are  compared with
 shell model results to investigate 
similarities  and differences between  two models
 in such observables like excitation energies and
M1 transitions  in $^{17}$C and  $^{21}$Ne. 

In this paper, we extend the previous calculations in ref.~\cite{Sagawa04}, 
 and particularly focus on recent 
experimental data of M1 transitions in  $^{17}$C and $^{21}$Ne 
 in order to study possible shape 
coexistence near the ground states of $^{17}$C.
This paper is organized as follows. 
We study the energy levels, magnetic moments  and 
   M1 transitions  by using  shell model wave functions  
in Section 2.  The deformed HF+blocked BCS results are shown in 
  Section 3.  
 A summary is given in Section 4.

\section{Shell model calculations of $^{17}$C and $^{21}$Ne}

\begin{table}
\vspace{-0.3cm}
 \squeezetable \caption{ \label{tab:Ex-shell}
 Shell model calculations of excitation energies  in $^{17}$C and $^{21}$Ne.
  The shell model calculations are performed by
using effective interactions PSDMK2, SFO and WBP.
 For sd shell configurations, the interaction matrices of 
  PSDMK2 and SFO are the same. Experimental data   
 are taken from ref.~\cite{RIKENC17} for $^{17}$C and from 
 ref.~\cite{Fire} for  $^{21}$Ne. All units are in MeV.} 
\begin{ruledtabular}
\begin{tabular}{c|c|c|c|c|c|c}
 &int.&$\frac{3}{2}_1^+$  &   $\frac{1}{2}_1^+$ & 
     $\frac{5}{2}_1^+$ &   $\frac{5}{2}_2^+$ &  $\frac{1}{2}_2^+$\\\hline
    $^{17}$C        & MK2  & 0.305 & 0.0 & 0.711 & 1.679  & 5.007\\ 
          &  SFO    & 0.304 & 0.0  & 0.654 & 1.678    & 5.039  \\  
          & WBP & 0.0   &  0.295  & 0.032 & 1.998 & 5.034  \\\hline
          &  exp   & 0.0  & (0.212)  &  0.333 &    &     \\\hline\hline
  & int.&$\frac{3}{2}_1^+$  &   $\frac{1}{2}_1^+$ & 
     $\frac{5}{2}_1^+$&   $\frac{5}{2}_2^+$ &  $\frac{1}{2}_2^+$ \\\hline
  $^{21}$Ne & MK2 (SFO) & 0.0 & 1.930 & 0.495 & 3.250 & 4.688 \\
            & WBP & 0.0 & 2.870 & 0.249& 3.484 & 5.815   \\\hline
          &  exp   & 0.0  & 2.794  &  0.351  &3.735  &     \\
\end{tabular}
\end{ruledtabular}
\end{table}

\begin{table}
\vspace{-0.3cm}
 \squeezetable \caption{ \label{tab:MM-shell}
 Magnetic moments  in $^{17}$C and $^{21}$Ne 
in unit of  $\mu_N^2$.
  The shell model calculations are performed by
using effective interactions PSDMK2, SFO and WBP with the bare g factors. The values in the brackets for $\frac{3}{2}_1^+$ in $^{17}$C and 
  $\frac{3}{2}_1^+$ and $\frac{5}{2}_1^+$ for $^{21}$Ne 
   are obtained by using 
the effective 
g factors for the IV channels, $\delta g_s=-0.2g_s^{IV}\tau_z,
\delta g_l=-0.15\tau_z$ and $g_p=-1.0\tau_z$ in Eq. (\ref{mu_eff}).
  Experimental data  of magnetic moments 
are taken from ref. \cite{C17mm} for  $^{17}$C and from ref. \cite{Rag89}
for $^{21}$Ne. } 
\begin{ruledtabular}
\begin{tabular}{c|c|c|c|c|c}
 $^{17}$C&$\frac{3}{2}_1^+$  &   $\frac{1}{2}_1^+$ & 
     $\frac{5}{2}_1^+$ &   $\frac{5}{2}_2^+$ &  $\frac{1}{2}_2^+$\\\hline
          MK2  & -0.710 (-0.686) & -1.548  & -1.453   & -1.447   & 0.280 \\ 
           SFO    & -0.725 (-0.713) & -1.548& -1.500   & -1.424   & 0.232\\
          WBP &  -0.858 ( -0.819) & -1.566 &  -1.404  &  -1.744   &0.570 \\\hline
           exp & $|0.758(38)|$  &  &  &  &        \\\hline\hline
      $^{21}$Ne &$\frac{3}{2}_1^+$  &   $\frac{1}{2}_1^+$ & 
     $\frac{5}{2}_1^+$ &   $\frac{5}{2}_2^+$ &  $\frac{1}{2}_2^+$\\\hline
  MK2 & -0.887 (-0.720)  & -1.498 &-0.657 (-0.484)  & -0.403  & 0.228\\
            WBP &  -0.824 (-0.674)   & -1.548 & -0.643 (-0.518) & -0.692  & 0.127\\\hline
           exp  &$ -0.661797(5)$ &   &   $|0.49(4)|$, & & \\
  & &   &   $|0.70(8)|$,$|0.88(20)|$   &  &   \\
\end{tabular}
\end{ruledtabular}
\end{table}

In light and medium mass nuclei, the shell model is one 
 of the most successful theories to describe nuclear
structure in both the ground states and the excited states.  
 Shell model calculations are performed in (p-sd) model space for
 $^{17}$C and (sd) model space for $^{21}$Ne with three 
effective interactions
 PSDMK2~\cite{MK2}, SFO~\cite{SFO} and WBP~\cite{WBP}.   
The excitation energies of the first $I^{\pi}$=
  3/2$^+$, 1/2$^+$ and 5/2$^+$ states are tabulated in
 Table~\ref{tab:Ex-shell}.  The SFO interaction is identical to PSDMK2 
 in (sd) model space so that two results  are the same for 
  $^{21}$Ne.  The excitation energies of  $^{21}$Ne are well reproduced 
 by all three shell model calculations.  It is not surprising since 
 the effective force is usually fitted to the data of stable nuclei such as 
   $^{21}$Ne.  The results of WBP show the best  agreement with
 experimental excitation energies within 100keV difference.
  The calculated results are much worse in the case of  $^{17}$C.  The 
 interactions PSDMK2 and SFO predict the spin-parity of the ground state 
 to be $I^{\pi}$=1/2$^{+}$, while the observed  spin-parity is 
 $I^{\pi}$=3/2$^{+}$.  The interaction WBP gives a state with 
  $I^{\pi}$=3/2$^{+}$ as the ground state.  However, 5/2$^{+}$ is almost 
 degenerate with the ground state contrary to the experimental data.

Magnetic moments and  magnetic dipole (M1)
  transition probabilities B(M1) are given in 
 Tables ~\ref{tab:MM-shell} and ~\ref{tab:M1-shell},respectively,
 The magnetic operator is defined as
\be
 {\bf \mu}_{eff}=(g_s^{bare}+\delta g_s){\bf s} +(g_l^{bare}+\delta g_l){\bf l}
    +g_p[Y_s\times {\bf s}]^{(1)}
\label{mu_eff}
\ee
where$\delta g_s$ and $\delta g_l$ are the renormalization factors for the spin and the orbital g factors, respectively.
  The last term of Eq. (\ref{mu_eff}) is the tensor component due to 
 the core polarization effect.  
  The shell model results of magnetic moments are shown in 
Table~\ref{tab:MM-shell} with the bare g factors and the effective 
g factors for the IV channels, $\delta g_s=-0.2g_s^{IV}\tau_z
=-0.2\frac{(g_s^{\nu}-g_s^{\pi})}{2}\tau_z, 
\delta g_l=-0.15\tau_z$ and $g_p=-1.0\tau_z$.
 For  the magnetic moments, the effect of $\delta g_s$ and $\delta g_l$ 
cancel each other largely and that of  the tensor component $g_p$ is 
very small.  The net effect of the effective operator is less than 
5\% in $^{17}$C and 20\% in $^{21}$Ne.  In comparison with experimental 
data,  the optimum  quenching factor $\delta g_s$ depends on the model 
space and the effective interaction.  For $^{17}$C, small quenching 
factors ($\delta g_s/g_s^{IV}\sim0.0$ for PSDMK2 and SFO, 
$\delta g_s/g_s^{IV}\sim-0.2\tau_z$ for WBP) give good agreement with the 
experimental data.  Slightly larger values ($\delta g_s/g_s^{IV}
\sim-0.25\tau_z$ for PSDMK2 and SFO, 
$\delta g_s/g_s^{IV}\sim-0.2\tau_z$ for WBP) give reasonable results in 
 the case of $^{21}$Ne.  

\begin{table}
\vspace{-0.3cm}
 \squeezetable \caption{ \label{tab:M1-shell}
 Shell model 
  B(M1) in $^{17}$C and $^{21}$Ne
in unit of  $\mu_N^2$ with  the bare g factors (the effective 
g factors).
  Experimental data are taken from ref.~\cite{RIKENC17} 
 for  $^{17}$C and ref.~\cite{Fire} for $^{21}$Ne.
See the captions to Table~\ref{tab:MM-shell} for details.   
 }
\begin{ruledtabular}
\begin{tabular}{c|c|c|c}
 &int.&$\frac{1}{2}^+_1\rightarrow\frac{3}{2}^+_1$ &
       $\frac{5}{2}^+_1\rightarrow\frac{3}{2}^+_1$ \\\hline
    $^{17}$C        & MK2  & 0.084 (0.045) & 0.070 (0.031)   \\
          &  SFO    & 0.077 (0.041) & 0.077 (0.035)\\
          & WBP & 0.078 (0.043)  &  0.077 (0.034) \\\hline
          &  exp  & 0.010$\pm$0.001  & 0.082 +0.032/$-$0.018        \\\hline\hline
     & int.&$\frac{1}{2}^+_1\rightarrow\frac{3}{2}^+_1$ &
                      $\frac{5}{2}^+_1\rightarrow\frac{3}{2}^+_1$ \\\hline
  $^{21}$Ne & MK2  & 0.724 (0.607) & 0.173 (0.128)   \\
            & WBP & 0.451 (0.390) &  0.161 (0.109) \\\hline
          &  exp  & 0.33$\pm$0.05 & 0.128$\pm$0.03        \\
\end{tabular}
\end{ruledtabular}
\vspace{-0.3cm}
\end{table}

In Table~\ref{tab:M1-shell},  two empirical 
M1 transition probabilities in  $^{21}$Ne are reasonably well 
 reproduced by the shell model calculations.  The best results among the 
three interaction are given by WBP interaction with the
effective spin g factor $\delta g_s/g_s^{IV}=-0.2\tau_z$.  
We can see in Tables ~\ref{tab:Ex-shell},~\ref{tab:MM-shell},~\ref{tab:M1-shell} that  the shell model 
provides good agreement not only for 
the excitation energies but also for the 
magnetic moments and M1 transition probabilities in  $^{21}$Ne. 
  In  $^{17}$C, the M1 
transition probability from  $I^{\pi}$=5/2$^{+}$ to 3/2$^{+}$ is 
 reproduced well by the shell model with the bare g factors.  
 However, the
 transition probability from  $I^{\pi}$=1/2$^{+}$ to 3/2$^{+}$ is 
 very poorly predicted, i.e., the empirical data is
 almost  one order magnitude 
  smaller than 
the shell model predictions with the bare g factors. The effective g
 factors adopted in the magnetic moments in Table~\ref{tab:MM-shell}
 decrease substantially the B(M1) values in $^{17}$C.  However, 
 these effective g factors do not give any satisfactory results for the 
measured two transitions between ($I^{\pi}=5/2^{+}\rightarrow 3/2^{+}$)
 and ($I^{\pi}=1/2^{+}\rightarrow 3/2^{+}$) as shown inside of the 
brackets in Table~\ref{tab:M1-shell}.  
Recently, the description of  M1 transitions in $^{17}$C has been
considerably improved with the use of a modified SFO Hamiltonian~\cite{SO}.

\section{Deformations and Magnetic Dipole Trabsitions 
in  $^{17}$C and  $^{21}$Ne}

  The neutron number dependence of
deformations was studied along the chain of  C and Ne isotopes in 
ref.~\cite{Sagawa04} by performing 
deformed HF+blocked BCS calculations with  Skyrme interactions 
SGII and SIII.  In this study, we perform the same 
deformed HF calculations of two N=11 isotones $^{17}$C 
and $^{21}$Ne  with a different Skyrme interaction SkO'.  
We found that the results of  SkO' are very close to those of
 SGII and SIII.  One advantage of  SkO' is to give an oblate deformed 
 ground state for $^{12}$C with the original spin$-$orbit interaction
, while the spin$-$orbit interaction was reduced  in SIII and SGII to obtain
 the oblate deformation.    
In numerical calculations, 
the axial symmetry is assumed  for  the HF deformed potential.
The pairing interaction  is taken to be a 
density dependent pairing interaction  in BCS approximation
For numerical details about the pairing calculations, 
see refs.~\cite{DHF,BRRM00}.  

\begin{figure}[hp]
\vspace{-0.3cm} 
\epsfig{file=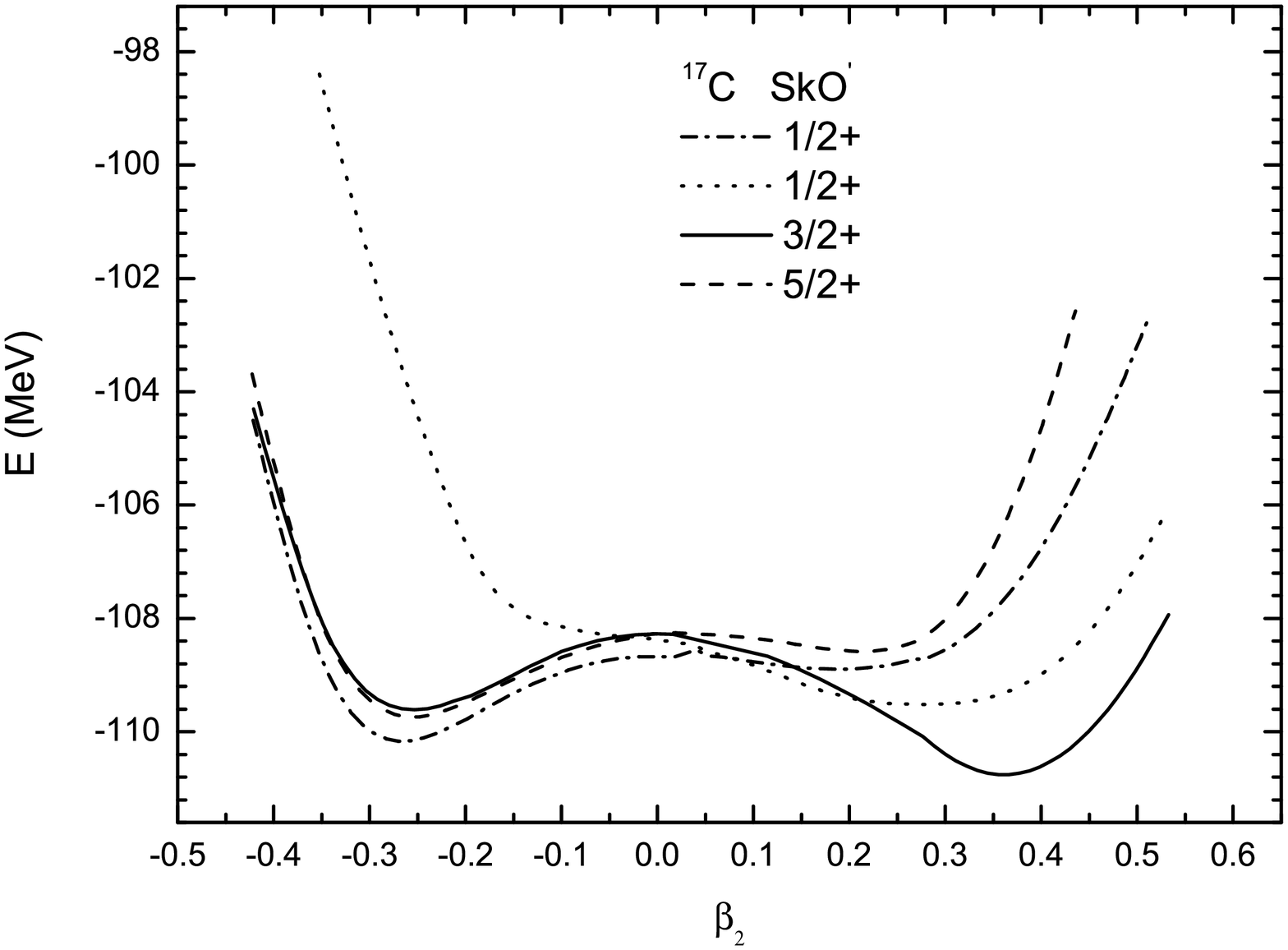,width=8cm}
\epsfig{file=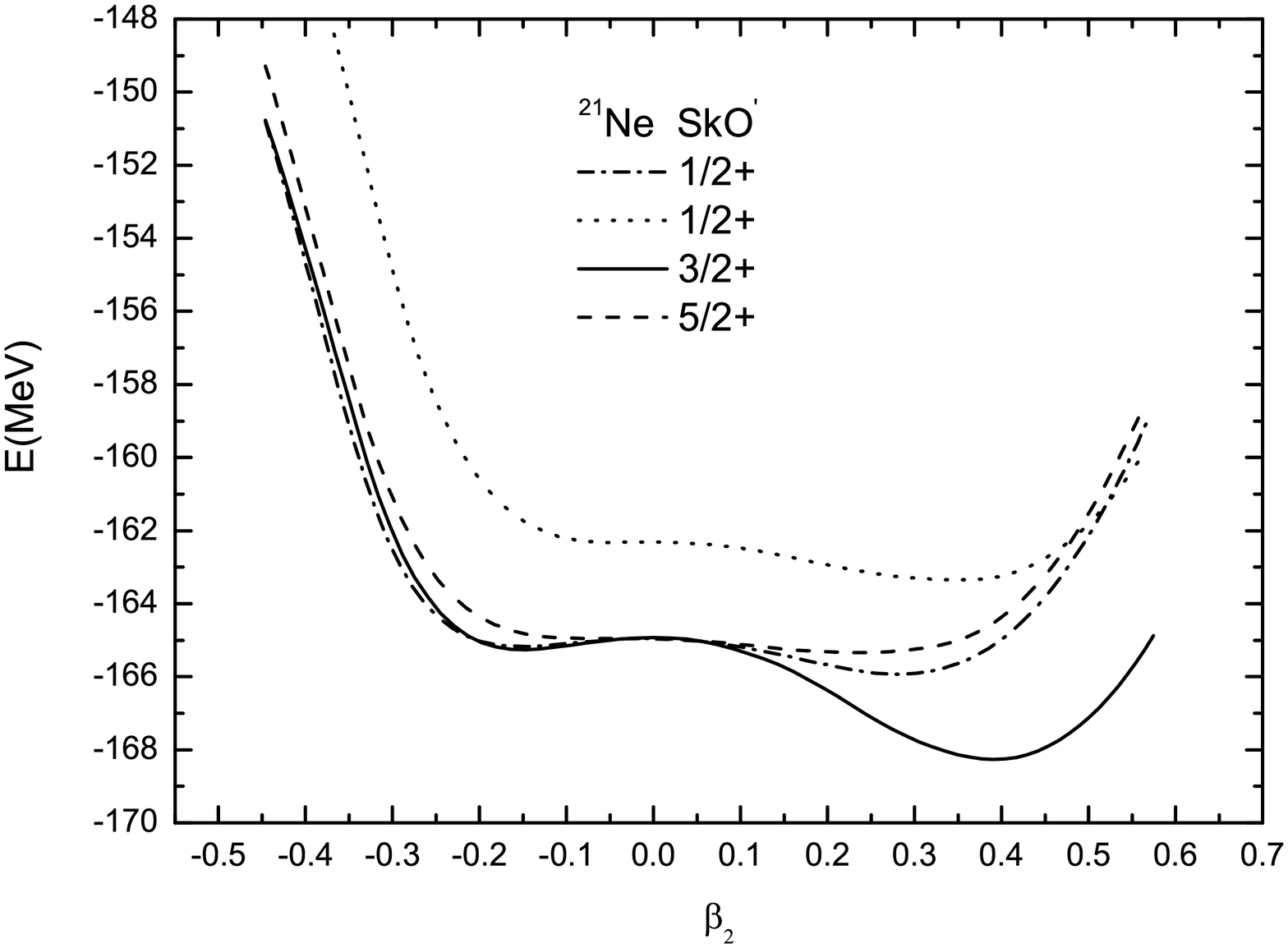,width=8cm}
\vspace{-0.3cm}
\caption{Energy surfaces as a function of deformation parameter $\beta_2$
 in  $^{17}$C and $^{21}$Ne. 
 Deformed HF+blocked BCS
 calculations are performed with a 
Skyrme interaction SkO'.}
\label{fig:oddHF}
\end{figure}

Deformed Skyrme HF+ blocked  BCS results   are shown in
 Fig.~\ref{fig:oddHF} 
(a) for $^{17}$C  and  Fig.~\ref{fig:oddHF} (b)
 for $^{21}$Ne.  
 The deformation and the 
intrinsic Q$_0$ moments are tabulated in Table
 \ref{tab:sko} for $^{17}$C, and  $^{21}$Ne.  
  The ground states are predicted to be $K^{\pi}=3/2^{+}$ state
  in both nuclei having large prolate deformations $\beta_2=$0.366 
for  $^{17}$C and 
 0.391 for $^{21}$Ne, respectively. The spin-parity of  calculated 
 results can be compared with  the observed ones  $I^{\pi}=3/2^{+}$ in both 
 nuclei.  
 In $^{17}$C, the first excited state is predicted to be  $K^{\pi}=1/2^{+}$
 state 
  with a large oblate deformation  $\beta_2=-$0.270.  The energy difference 
 from the ground state is rather small with  E$_{\mbox{x}}$=0.56MeV. 
  On the other hand, the first excited $K^{\pi}=1/2^{+}$ in  $^{21}$Ne 
  is predicted to have
  a large prolate deformation  $\beta_2=$0.287 with   a large 
  excitation energy E$_{\mbox{x}}$=2.33MeV. 
  This difference in $K^{\pi}=1/2^{+}$ state can be understood
  as a manifestation of 
the nuclear Jahn-Teller effect due to the  proton configuration
  \cite{NJT}.
  In general, a few particles top of the closed shell drives prolate deformation, while a few  holes prefer oblate deformation.
  There is a strong competition between prolate driving N=11 neutrons and
 oblate driving Z=6 protons in  $^{17}$C.  
  Namely, the Z=6 proton configuration, two proton holes in the 
Z=8 closed shell,  prefers the oblate 
  deformation as is the case of the ground state of $^{12}$C, while 
   the N=11 neutrons tends to drive prolate deformation.  
Consequently, in   $^{17}$C, the ground state is prolately  deformed 
  due to the effect of neutron configuration.  However, the first excited 
   $K^{\pi}=1/2^{+}$ state becomes  oblate under the influence of
  the deformation driving force
  of  protons.  In $^{21}$Ne, both the proton and neutron configurations  
  drive prolate deformation so that there is 
  no sign of the shape coexistence.
 The observed excitation energy of the first  $I^{\pi}=1/2^{+}$ state is 
 very low in  $^{17}$C as  E$_{\mbox{x}}$=0.212MeV, while that of 
  $^{21}$Ne is higher as  E$_{\mbox{x}}$=2.79MeV. These 
  experimental observations are 
 consistent with the calculated results
   in Table \ref{tab:sko} as far as the 
 excitation energies are concerned.  Thus we identify the first excited 
$I^{\pi}=1/2^{+}$ state 
  as  $K^{\pi}=1/2^{+}$ in both  $^{17}$C and  $^{21}$Ne with different 
  large   deformations   $\beta_2=-0.270$ and 0.287, 
  respectively. The $I=1/2^{+}$ in  $^{21}$Ne was interpreted  in 
  \cite{Rolfs72}   as the  head of 
 rotational band with a large prolate deformation \cite{Rolfs72}. 
 The  $I^{\pi}=5/2^{+}$ state is observed at very 
low excitation energy around  E$_{\mbox{x}}$=0.3MeV in both nuclei.
 In the HF calculations, 
no  $K^{\pi}=5/2^{+}$ state appears at the energy below  E$_{\mbox{x}}\sim$
1MeV.  Thus, we interpret that the observed first excited $I^{\pi}=5/2^{+}$ 
state in both nuclei is a  member of the rotational band with  
  $K^{\pi}=3/2^{+}$.  In the case of  $^{21}$Ne, the ground state and the 
 first excited state were identified as members of the same rotational band 
   giving consistent predictions  
of associated observed properties \cite{Rolfs71}.   This 
interpretation is also supported by the large deformation 
 length observed in the excitation to $I^{\pi}=5/2^{+}$ state in the proton 
 inelastic scattering on  $^{17}$C \cite{Elekes05}.

\begin{table}
\vspace{-0.3cm}
 \squeezetable
  \caption{\label{tab:sko}
Energies, deformations, $Q$ moments and magnetic moments 
in $^{17}$C and  $^{21}$Ne
with a Skyrme interaction SkO'. The magnetic moment $\mu$ is calculated for
  $I=K$ state with the bare neutron g$-$factor. 
   Experimental data are  the same as for Table~\ref{tab:MM-shell} 
(experimental uncertainties are omitted).}

\begin{ruledtabular}
\begin{tabular}{l|r|r|r|r|r|r|r|c}
 &$K^{\pi}$& $E_x$&$\beta_2$& $Q_{0p}$ & $Q_{0n}$&$g_k$&$\mu$&$\mu$(exp)\\
         &         & &            &     (fm$^2$)    & (fm$^2$)  &&$(\mu_N$)  &  $(\mu_N$)  \\
\hline \hline
$^{17}$C&$\frac{3}{2}^+_1$&0.0 & 0.366& 16.24& 53.05&-1.197&-0.877& $|0.758|$ \\
        &$\frac{1}{2}^+_1$&0.56&-0.270&-15.27&-35.94&-3.420&-1.767&\\
        &$\frac{5}{2}^+_1$&0.99&-0.247&-13.40&-34.43&-0.764&-1.126&\\
        &$\frac{1}{2}^+_2$&1.21& 0.272& 13.59& 40.75&-1.101&-0.947 & \\ \hline \hline
$^{21}$Ne&$\frac{3}{2}^+_1$&0.0& 0.391&  42.15& 46.98&-1.112&-0.728& -0.661797\\
         &$\frac{1}{2}^+_1$&2.33&0.287&31.67&32.05&-2.557&-1.523& \\
         &$\frac{5}{2}^+_1$&2.92&0.226&25.85&23.83&-0.764&-1.040& $|0.49|$
 $\sim$  
\\ &&&&&&&&  $|0.88|$\\
&$\frac{1}{2}^+_2$&4.91&0.357&39.570&43.547&-0.231&-0.594& \\
\end{tabular}
\end{ruledtabular}
\end{table}

\begin{table}
\vspace{-0.3cm}
 \squeezetable \caption{ \label{tab:M1-def}
 M1 transition probabilities B(M1) in $^{17}$C and $^{21}$Ne 
in unit of  $\mu_N^2$.
  The deformed HF calculations are  performed by
using a Skyrme interaction SkO'.  The effective spin g-factor
  $g_s^{eff}=0.5g_s^{bare}(0.7g_s^{bare})$ is adopted. 
Experimental data   are taken from ref.~\cite{RIKENC17} for $^{17}$C and
 from  ref.~\cite{Fire} for  $^{21}$Ne.  
 } 
\begin{ruledtabular}
\begin{tabular}{c|c|c}
 &$I_1(K)=\frac{1}{2}^+_1\rightarrow I_2(K')=\frac{3}{2}^+_1$ &
       $K=\frac{3}{2}, I_1=\frac{5}{2}^+_1\rightarrow I_2=\frac{3}{2}^+_1$ \\\hline
    $^{17}$C    & 0.00068(0.00094) & 0.116(0.179)  \\ \hline
         exp  & 0.010$\pm$0.001 & 0.082 +0.032/$-$0.018        \\\hline\hline
   $^{21}$Ne  & 0.208(0.272) & 0.152(0.224)  \\\hline
        exp  & 0.33$\pm$0.05 & 0.128$\pm$0.003        \\
\end{tabular}
\end{ruledtabular}
\end{table}

We study the  magnetic dipole transitions between  the excited and ground 
 states 
in $^{17}$C and $^{21}$Ne  using 
  the deformed HF   wave functions.  

For axially symmetric deformation, 
the deformed many-particle initial and final 
  states are expressed as 
 a  direct product of neutron and proton single-particle states;
\begin{align}
|K\rangle &= |\nu\rangle |\pi\rangle ,
\label{eq:many-qp}
\end{align}
where the component of the total angular momentum along the symmetry axis
is denoted by $K$~\cite{BM2,Nilsson55} and 
$|\nu (\pi) \rangle = a^{\dagger}_{\rho_{1}} a^{\dagger}_{\rho_{2}} \cdots 
 |f (\beta_2)\rangle$   
denotes the multi-quasiparticle neutron (proton) 
 state.     
The state $|f (\beta_2)\rangle$ is the
quasiparticle vacuum  with deformation $\beta_2$.
 The quasiparticle operator  $a^{\dagger}$ 
 is connected to the real particle operators 
 $c^{\dagger}(\beta_2)$
and  $c(\beta_2)$
in the deformed  basis by
\begin{equation}
a^{\dagger}_{\lambda\mu}(\beta_2) = u_{\lambda\mu}(\beta_2) c^{\dagger}_{\lambda\mu}(\beta_2)
- v_{\lambda\mu}(\beta_2) c_{\widetilde{\lambda\mu}}(\beta_2) ,
\end{equation}
where $\lambda$ specifies the quantum numbers of Nilsson orbit,   
 $v_{\lambda\mu}(\beta_2)$ is the  BCS occupation amplitude and 
 $u_{\lambda\mu}(\beta_2)=\sqrt{1-v_{\lambda\mu}(\beta_2)^2}$. The 
 operators $ c^{\dagger}_{\lambda\mu}(\beta_2)$ and 
 $ c_{\lambda\mu}(\beta_2)$ are further expanded by the spherical bases as 
$c^{\dagger}_{\lambda\mu}(\beta_2) = \sum_{a} d^{\mu}_{\lambda a}(\beta_2) c^{\dagger}_{a\mu}$  
where the amplitude $d^{\mu}_{\lambda a}(\beta_2)$ is denoted by the 
quantum numbers $a=(n,l,j)$.

The intrinsic M1 single-particle 
transition operator is expressed as
\beq
 &{\mathcal M}(M1)= \sqrt{\frac{3}{4\pi}}\mu_N \times \nonumber \\  
 &\left(\sum_i
\left( \left(g_l(i)-g_R\right){\bf l_i} 
 + \left(g_s(i)-g_R\right){\bf s_i} \right)+g_R{\bf I}  \right)
\label{eq:M1}
\eeq
where $g_l(i)$, $g_s(i)$ and $g_R$ are the orbital, spin $g$ factors and
the gyromagnetic ratio of the rotor 
 respectively, in unit of the nuclear magneton $\mu_N=e\hbar/2m_pc$.  

The  transition matrix element can be written for one neutron quasiparticle 
states as 
\beq
&\langle \nu' \pi'  K' |{\mathcal M}(M1)
| \nu \pi K \rangle  =  \nonumber \\
&\langle \nu' |a_{\lambda' K'}(\beta_2')
 {\mathcal M}(M1)_n a^{\dagger}_{\lambda K}(\beta_2) | \nu \rangle
\langle \pi' | \pi \rangle
\eeq
where $\langle \pi' | \pi \rangle$ and
 $ \langle \nu' | \nu \rangle$ are quasi-particle vacuum overlaps of neutrons and protons, respectively.

The in-band  M1 transition probability can
be written, for a band with $K>\frac{1}{2}$,  as
\beq
&B(M1;KI_1\rightarrow K,I_2=I_1\pm1)= \nonumber \\
&\frac{3}{4\pi} \mu_N^2(g_K-g_R)^2K^2
               <I_1K10|I_2K>^2
\label{eq:mu1}, 
\eeq 
where 
$g_R=\frac{Z}{A}$ 
and $g_K$ is the intrinsic $g$ factor,
$Kg_K= \langle K|g_ll_3+g_ss_3 |K\rangle$.  
The magnetic moment is expressed as 
\be
 \mu=g_RI+(g_K-g_R)\frac{K^2}{I+1}.
\ee
For $K\neq K'$ case, the M1 transition probability is written to be
\beq
& B(M1;KI_1\rightarrow K',I_2)= \frac{3}{4\pi} \mu_N^2 \times \nonumber \\
&  <I_1K1K'-K|I_2K'>^2 
 G^2 <\pi'|\pi>^2
\eeq
where
\beq
G=  \left(g_s-g_R\right)\langle \nu' |a_{\lambda' K'}(\beta_2')
s_{\Delta K} a^{\dagger}_{\lambda K}(\beta_2) | \nu \rangle
 \nonumber \\
+\left(g_l-g_R\right)\langle \nu' |a_{\lambda' K'}(\beta_2') l_{\Delta K}
  a^{\dagger}_{\lambda K}(\beta_2) | \nu \rangle .
\eeq
with $\Delta K=K'-K$.  

The calculated B(M1) values are tabulated in Table \ref{tab:M1-def}.
 We adopt $g_s^{eff}=0.5g_s^{bare}$
and  $g_s^{eff}=0.7g_s^{bare}$
  for the effective neutron spin g-factor in the calculations.
One can see a large hindrance in  B(M1) from $I_1=1/2^+$ to $I_1=3/2^+$ 
  transition in  $^{17}$C.  This is entirely due to the shape 
difference 
  between the ground  and the first excited states. Namely,  
the value of the core overlap of BCS vacuums  
$<q'|q>\equiv<\nu'|\nu><\pi'|\pi>$  is calculated to be
 $<q'(\beta_2=0.366)|q(\beta_2=-0.270)>$=0.0378
 between $I=3/2^+_1$ and  $I=1/2^+_1$ states in $^{17}$C. On the other
 hand the corresponding overlap in $^{21}$Ne is close to 1.0, i.e.,  
 $<q'(\beta_2=0.391)|q(\beta_2=0.287)>$=0.982 since both 
the initial and the final 
 state have large prolate deformations.
  For in-band transition ($I_1=5/2^+\rightarrow I_2=3/2^+,K=3/2)$, 
 the calculated  B(M1) values with the g-factor $g_s^{eff}=0.5g_s^{bare}$ 
agree well with the observed ones 
  within the experimental accuracies.  This quenching factor is somewhat smaller 
 than the adopted values in rare-earth nuclei, but it is  still in the 
acceptable range.  
  The g-factor $g_s^{eff}=0.7g_s^{bare}$ gives better results for 
 the transition ($I_1=K_1=1/2^+\rightarrow I_2=K_2=3/2^+)$
   in $^{21}$Ne.
   The deformed HF calculations gives  the lowest $K_1^{\pi}=1/2^+$ state 
having  asymptotic quantum numbers
 $ [Nn_3\Lambda\Omega]=[2201/2]$.  There is another $K^{\pi}=1/2^+$ state 
 with  $ [Nn_3\Lambda\Omega]=[2111/2]$ having slightly higher energy.
  The latter has about 2 times larger $B(M1)$ value for the transition to
the $K=3/2$ ground state. It is expected that 
 a  small mixing of the $ [Nn_3\Lambda\Omega]=[2111/2]$ state  increases
  the $B(M1)$ value between 
the ($I_1=K_1=1/2^+\rightarrow I_2=K_2=3/2^+)$.  
We notice that the optimal quenching spin g factor of the deformed HF 
results is slightly smaller than that of the shell model calculations.
 It is an interesting open  question to compare  more systematically 
the transition strength of two models in a quantitative level.  
It was mentioned in ref.~\cite{RIKENC17} that the halo effect of $^{17}$C may play a role to decrease B(M1) value of 
($I_1=1/2^+\rightarrow I_2=3/2^+)$. However no serious attempt 
 has been made so far to take into account the halo effect on the M1 
 transitions in $^{17}$C.  

The calculated magnetic moments $\mu$ are shown in Table  \ref{tab:sko}.
 The observed magnetic moments show a small quenching effect 
in comparison 
with the calculated values for   
the ground states of $^{17}$C and $^{21}$Ne. The results of deformed 
HF calculations provide similar quantitative predictions to those of 
the shell models as far as the magnetic moments of the ground states
  are concerned. 
The observed magnetic moment for the excited $I^{\pi}=5/2^+$ state
  in  $^{21}$Ne is still not accurate enough to perform 
precise comparison with the calculated results.  

The second $K=1/2^+$ state is found in Table \ref{tab:sko} 
at rather low energy $E_x=$1.21MeV $^{17}$C by  the
deformed HF model, while the $I=1/2_2^+$ state is located at $E_x\sim$5MeV 
in the shell model calculations in Table~\ref{tab:Ex-shell}. 
So far the second $1/2^+$ is not identified experimentally \cite{Boh07}.
It is quite interesting to find the $1/2_2^+$ state experimentally 
  to disentangle the applicability of the two models.  

\section{SUMMARY}
 We have studied the magnetic dipole transitions in $^{17}$C 
and $^{21}$Ne using microscopic shell 
model wave functions and deformed HF wave functions.
The energy spectra as well as M1 transition probabilities of  $^{21}$Ne
 are well 
reproduced  by the shell model calculations,  while we  need a 
  quenching factor for  the spin g factor to obtain  reasonable 
quantitative agreement.
 On the other hand, the observed M1 transition probability
  from the first excited 
 1/2$^+$ to the 3/2$^+$ ground state  in  $^{17}$C was found 
to be  hindered by one order 
of magnitude compared with the shell model calculations. 
The shell model prediction of energy spectra is also poor in  $^{17}$C
 compared with the experimental data.  
The deformed HF+blocked BCS calculations are performed with a 
Skyrme interaction SkO'.  The ground states of $^{17}$C and $^{21}$Ne are 
  predicted as  largely prolate deformed states with $K^{\pi}=3/2^+$.
In $^{21}$Ne, the first $K^{\pi}=1/2^+$ state appears at the energy 
 $E_{\mbox{x}}\sim$2.3MeV  with a large prolate deformation. 
On the other hand, the first $K^{\pi}=1/2^+$ state in $^{17}$C has a large 
oblate deformation with $E_{\mbox{x}}\sim$0.5MeV as the result of competition between the deformation driving force of protons and neutrons.   
The calculated energy difference between  $K^{\pi}=3/2^+$ and  
 $K^{\pi}=1/2^+$ states is close to the observed energy difference between 
  $I^{\pi}=3/2^+$ and   $I^{\pi}=1/2^+$ states both
 in  $^{21}$Ne and $^{17}$C.
 The strong hindrance of the B(M1) transition between the first excited 
 1/2$^+$ to the ground state 3/2$^+$ of $^{17}$C  
can be attributed to the 
  shape difference between the lowest $K^{\pi}=1/2^+$ 
  and the first $K^{\pi}=3/2^+$ state  as is  
 predicted by the deformed HF+blocked BCS calculations.

We thanks H. Sakurai for communications on experimental data prior to
publication.
This work is 
supported in part by the Japanese  
Ministry of Education, Culture 
,Sports, Science  and Technology 
  by Grant-in-Aid  
for Scientific Research under
 the program number C(2) 18540290, 20540277,  
  the National Science Foundation of China
under contract No.~10605018 and the program for New Century Excellent 
  Talents in University 
under contract No.~NCET-07-0730.


%

%
%
\newpage

\begin{thebibliography}{99}
\bibitem{RIKENC17}
  D. Suzuki et al., Phys. Lett. B666, 222(2008).
\bibitem{Fire}
{\it Table of Isotopes}, eds. by R. B. Firestone et al.,
(Wiley, New York, 1996). 
\bibitem{Sagawa04} H. Sagawa, X. R. Zhou and X. Z. Zhang,
 Phys. Rev. C{\bf 70}, 054316(2004).
\bibitem{MK2}
 D. J. Millener and D. Kurath, Nucl. Phys. {\bf A255}, 315 (1975). 
\bibitem{SFO}
 T. Suzuki, R. Fujimoto, T. Otsuka, Phys. Rev. C {\bf 67}, 044302 (2003).
\bibitem{WBP}
E. K. Warburton and B. A. Brown, Phys. Rev. C{\bf 46}, 923 (1992);\\
OXBASH, the Oxford, Buenos-Aires, Michigan State, Shell Model Program,
B. A. Brown et al., MSU Cyclotron Laboratory
Report No. 524, 1986.
\bibitem{SO}
T. Suzuki and T. Otsuka, Proceedings of ISPUN07 (Hoi An, Vietnam, 2007, World 
Scientific), p.3138 and to be published.
\bibitem{DHF}
H. Sagawa, T. Suzuki and K. Hagino, Nucl. Phys. A{\bf 722}, 183 (2003).
\bibitem{BRRM00}M. Bender, K. Rutz, P.-G. Reinhard, and J.A. Maruhn,
Eur. Phys. J. {\bf A8}, 59 (2000). 
\bibitem{NJT}
  H. A. Jahn and E. Teller, Proc. R. Soc. London, Ser. A {\bf 161}, 
  220(1937). \\
P.-G. Reinhard and E. W. Otten, Nucl. Phys. {\bf A420}, 
173 (1984).\\
 W. Nazarewicz, Int. J. Mod. Phys. E{\bf2}, 51 (1993); Nucl. Phys. 
  {\bf A574}, 27c (1994). 
\bibitem{Rolfs72}
  C. Rolfs et al., Nucl. Phys. A189, 641(1972).
\bibitem{Rolfs71}
 C. Rolfs et al., Nucl. Phys. A167, 449(1971).
\bibitem{Elekes05}
  Z. Elekes et al., Phys. Lett.  B614, 174(2005).
\bibitem{BM2}
 A. Bohr and B. R. Mottelson, Nuclear Structure Vol. 2 (1975, 
  W. A. Benjamin, Inc.)
\bibitem{Nilsson55}
  S. G. Nilsson, Mat. Fys. Medd. Dan. Vid. Selsk. {\bf 29}, no. 16 (1955).

\bibitem{C17mm}
H. Ogawa et al.,
Euro. Phys. J.  {\bf A13}, 81 (2002). 
\bibitem{Rag89}
P. Raghaven, Atomic Data Nucl. Data Tables {\bf 42}, 189 (1989).
\bibitem{Boh07}
  H. G. Bohlen et al., Eur. Phys. J. A{\bf 31}, 279 (2007).   
\end{thebibliography}
\end{document}